# Estimation Enhancing in Optoelectronic Property: A Novel Approach Using Orbital Interaction Parameters and Tight-Binding

Ali Haji Ebrahim Zargar +(Iran University of Science and Technology, University St., Hengam St., Resalat Square, Tehran, Iran)   (alizargar1372@gmail.com)

Ali Amini ( Amirkabir University of Technology, Hafez Ave, Valiasr Square, Tehran, Iran) (ali.amini.scholar@gmail.com)

Ahmad Ayatollahi (Iran University of Science and Technology, University St., Hengam St., Resalat Square, Tehran, Iran)   (ayatollahi@iust.ac.ir)

## Abstract

This paper advocates for an innovative approach designed for estimating optoelectronic properties of quantum structures utilizing Tight-Binding (TB) theory. Predicated on the comparative analysis between estimated and actual properties, the study strives to validate the efficacy of this proposed technique; focusing notably on the computation of bandgap energy. It is observed that preceding methodologies offered a restricted accuracy when predicting complex structures like super-lattices and quantum wells. To address this gap, we propose a methodology involving three distinct phases using orbital interaction parameters (OIPs) and the TB theory. The research employed Aluminium Arsenide (*AlAs*) and Gallium Arsenide (*GaAs*) as the primary bulk materials. Our novel approach introduces a computation framework that first focuses on bulk computation, subsequently expanding to super-lattice structures. The findings of this research demonstrate promising results regarding the accuracy of predicated optoelectronic properties, particularly the cut-off wavelength. This study paves the way for future research, potentially enhancing the precision of the proposed methodology and its application scope within the field of quantum optoelectronics.

**Keywords: orbital interaction parameter (OIP)_ Tight-Binding (TB)_ Genetic algorithm (GA)**

## Introduction

The current interest in quantum structures such as two-dimensional super-lattices and quantum wells, grown using Molecular Beam Epitaxy (MBE) and Metalorganic Chemical Vapor Deposition (MOCVD), is driven by their unique optoelectronic characteristics [1–3]. These characteristics are highly sensitive to variations in structure thickness due to quantum confinement. Therefore, any changes in the thickness of these quantum structures could lead to modifications in the transmission and optoelectronic properties of the resultant material. It is standard practice to evaluate the optoelectronic and electrical properties of these quantum structures post-growth. The aim of this process is to ascertain whether the grown structure meets the designer's specifications. However, due to the high costs associated with growth and characterization, both in terms of finance and time, it is more practical to simulate and predict the optoelectronic properties of the proposed structure prior to growth. If the simulations yield positive results, then the growth process can proceed. Therefore, a system that can accurately predict the properties of the material is of paramount importance to material designers and engineers.

Multiple simulation methodologies have been developed for quantum structures, one of which is the Density Functional Theory (DFT). DFT has proven to be a reliable method for predicting optoelectronic properties [4, 5], and for determining the ground state properties of materials. However, it requires the inclusion of quasi-particle corrections [6] and the excitonic effect [7]. Moreover, DFT has been shown to lack considerable accuracy for bandgap energy calculations, which are crucial in semiconductor-based devices [8], and it is not a viable alternative for structural calculations in systems with a large number of atoms[9].

Recently, alternative methods based on TB have been developed, which are capable of predicting optoelectronic properties with great accuracy and speed [10–12]. Here, the number of neighbors and the orbitals included in the computations have a significant impact on the accuracy of the predictions of the quantum and transfer properties. The computations and formation of the Hamiltonian matrix are influenced by each orbital and their associated OIPs. Therefore, accurate parameter estimation is crucial in predicting these properties. To increase the accuracy of these methods, phenomena such as segregation, which occur during the epitaxial growth of the structure, can be considered in the analyses [13,14].

Numerous strategies have been proposed to optimize OIPs, particularly for materials in the III-V group [15-19]. A notable contribution to this field was made by Vogl et al. [18], who established mathematical relationships between OIPs by diagonalizing the Hamiltonian matrix of the bulk structure. Utilizing a least-squares optimization method, they optimized the OIPs to predict optoelectronic properties closely aligned with the experimental values found in the band structure of group III-V compounds. However, their study only considered the bandgap energy levels for estimation and did not consider the effective mass, thus overlooking the electron transfer properties affected by variations in the graph's slope. Moreover, their model inadequately represented the split-off bands [18].

Subsequently, the Hamiltonian matrix was expanded by Klimeck et al. [17,20] using the sp3s* approximation. They undertook a different approach by optimizing the OIPs in bulk III-V materials using a genetic algorithm on the band structure, aiming to predict both the energy values and the effective mass of group III-V compounds. Notably, unlike Vogl, Klimeck did not use any mathematical relationships to optimize the OIPs. Despite this, the parameters were successful in accurately predicting the experimental properties of bulk group III-V compounds.

Machine Learning (ML) and Deep Learning (DL) techniques have been progressively utilized to delve into the electronic and optoelectronic characteristics of quantum structures. These advanced ML and DL algorithms significantly boost traditional computational methods such as TB and DFT. For example, [22] successfully utilized DL to make predictions about the electronic properties of a wide scope of two-dimensional materials, while [23] implemented ML to predict band gaps of double perovskites, an integral data for quantum structure research. These prominent studies assert that ML and DL have fundamentally transformed our comprehension and forecasts of the properties of quantum structures, thereby providing a potent and promising apparatus for future studies in this field. Additionally, recent studies like [23,24] have prognosticated the optoelectronic properties of materials through the estimation of the DFT diagram of the proposed material using the bulk structure of the target material to optimize the OIPs. This method amalgamates the calculation speed of TB with the precision of DFT. Moreover, the estimation points are not restricted to just experimental data but can extend to additional points using the DFT approach. Nevertheless, the reliability of results from the DFT approach in determining the optoelectronic properties stands relatively less dependable.

This paper will introduce an innovative methodology for computing OIPs that not only assures precise accuracy in bulk property estimation computations, but also predicts and estimates the properties of quantum structures like super-lattices and quantum wells with minimal error. Chapter two will delineate the proposed strategy for optimizing and calculating OIPs. In the subsequent chapter, we will apply this method to the $(GaAs)_m/(AlAs)_n$ super-lattice and the $GaAs/Al_xGa_{1-x}As$ quantum well structures. The objective is to compare the estimated and actual property values, thereby verifying the effectiveness of our proposed method. Specifically, we will scrutinize the accuracy of calculating the bandgap energy, a key indicator of the optoelectronic characteristics of the material, using alternative methodologies and their experimental results. The final chapter will provide a summarization of the findings and offer insights into potential areas for future research.

## Proposed Method

As delineated in the previous sections, the optoelectronic properties of quantum structures can be approximated using the OIPs and the TB methodology. However, it was observed that previous research primarily focused on optimizing OIPs for estimating optoelectronic properties in the bulk of quantum structures, thereby limiting the precision of this method for predicting the properties of more complex structures such as super-lattices and quantum wells. To address this limitation, this section introduces a novel technique for optimizing the OIPs to accurately estimate and predict the electrical and optoelectronic characteristics of not only the bulk but also other quantum structures, including super-lattices and quantum wells. In the proposed method, the TB methodology along with the sp3s* approximation are employed to construct the Hamiltonian matrix utilizing the OIPs. This approach takes into consideration both the first nearest neighbors and the spin-orbit effect, which are crucial for accurate computation of the optoelectronic characteristics post-OIP optimization.

In the empirical tight-binding method utilizing the sp3s* approximation, there are fifteen OIPs. Among these, six parameters represent the energies of s, p, and s* orbitals for both anion and cation components in a material. Seven parameters are dedicated to the interactions between atomic orbitals, capturing the electron hopping probabilities across different types of orbitals. The remaining two parameters pertain to spin-orbit interactions, which are essential for understanding certain material properties. For example, a parameter like $E_{xayc}$ signifies the interaction energy between the $P_x$ orbital in the anion and the $P_y$ orbital in the cation.

The fundamental layout of this method for OIP optimization is represented in Figure 1. This diagram provides a graphical representation of the process, from the formulation of the Hamiltonian matrix to the computation of optoelectronic properties.

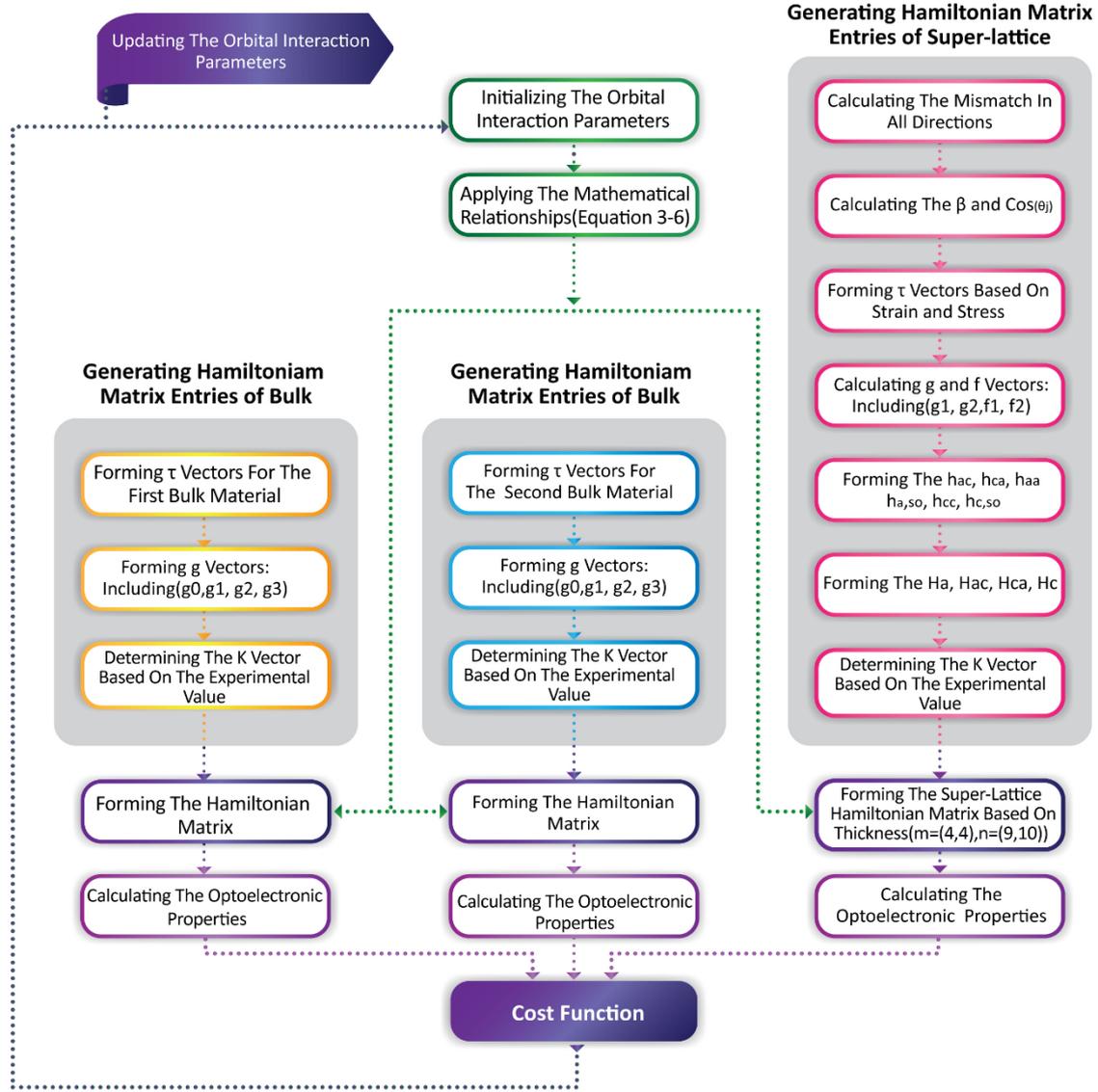

Figure 1: Diagram of the proposed method for optimizing the OIPs using a novel cost function and genetic algorithm

As depicted in Figure 1, our research methodology encompasses three distinct phases, each contributing to the formulation of the cost function. The first two phases involve calculations performed on bulk materials. In this study, we focus on *AlAs* and *GaAs*, which serve as our bulk materials. For each of these bulks, we generate vectors τ and g. Subsequently, the K vector is computed, which is associated with the material property under investigation. We then construct the Hamiltonian matrix for the material in question, leveraging OIPs and established mathematical relationships. The final step in this phase is the computation of the required optoelectronic property for the appropriate bulk material. The third phase involves a process similar to that performed for the bulk materials, but this time applied to the $(GaAs)_m/(AlAs)_n$ super-lattice, where $m$ equals 9 or 10, and $n$ equals 4.

In this phase, we calculate the mismatch in all directions, and subsequently compute β and $\cos(\theta_j)$. After generating τ, g, and $f$ vectors, we may obtain $h_{ac}$, $h_{ca}$, $h_{aa}$, $h_{a,so}$, $h_{cc}$ and $h_{c,so}$. We then form the matrices $H_a$, $H_{ac}$, $H_{ca}$, and $H_c$. The Hamiltonian matrix for the super-lattice is computed using the OIPs, and the K vector is determined based on the optoelectronic property of interest and specific points in the first Brillouin zone where experimental values for the optoelectronic property are reported.

As previously discussed and detailed in the appendix of [19], constructing the Hamiltonian matrix in this context involves initially creating several block matrices. Each block matrix comprises interior matrices that define the interactions between anion and cation materials within the Hamiltonian framework. The primary matrices in this construction are $h_{ac}$ and $h_{ca}$. These matrices are composed of elements including coordinate-specific variables ($\tau$), OIPs, factors $g$ and $f$, the reciprocal lattice vector ($K$), and terms like $\beta$ and $\cos(\theta_j)$. These specific terms are derived from the geometrical configuration of the super-lattice, considering the effects of strain and stress. Furthermore, the appendix elaborates on the corresponding $g$ and $\tau$ vectors, as well as the methodology behind the formation of the Hamiltonian matrix in bulk materials.

After performing the necessary computations for the bulk and super-lattice, we use the cost function corresponding to equation (1) to estimate the optoelectronic properties. The OIPs are optimized using the Genetic Algorithm (GA), a popular evolutionary metaheuristic algorithm. In this algorithm, we initially generate a population of 10,000 OIPs randomly. We also implement constraint-dependent crossover and mutation, and the GA executes 453 iterations before the optimization process is terminated.

$$Cost\ Function = \sum_{i=1}^{N=23} \lambda_{1,i}|X_i - X_i'| + \lambda_{2,i}|Y_i - Y_i'| + \lambda_3|W - W'| + \lambda_4|Z - Z'| \quad (1)$$

$$X_i, Y_i, W, Z = Function(E_{sa}, E_{sc}, E_{ssa}, E_{ssc}, E_{xayc}, E_{saxc}, E_{xasc}, E_{ssaxc}, E_{xassc}, E_{pa}, E_{pc}, E_{sasc}, E_{xaxc}, \Delta a, \Delta c) \quad (2)$$

In Equation (1), $X_i$ and $Y_i$ represent the calculated values of specific characteristics, while $X_i'$ and $Y_i'$ stand for the experimental figures of optoelectronic properties documented for the bulk of *AlAs* and *GaAs*. $Z'$ and $W'$ denote the experimental values of the optoelectronic characteristics. The calculated values of the optoelectronic property within the super-lattice quantum structure are symbolized by $Z$ and $W$. Equation (2) further elucidates the encompassing relationship between the OIPs and optoelectronic properties.

The $\lambda$ coefficients are utilized as hyper-parameters to adjust the cost function. The selection of these hyper-parameters is contingent upon the significance of estimating each of the optoelectronic properties. The $\lambda$ coefficients for the first and second components of the cost function are computed distinctively, depending upon the type of estimable optoelectronic properties. Given the crucial nature of precise bandgap energy estimation in this study, the values corresponding to this property in the bulk are designated as 100,000. For other bulk optoelectronic properties, $\lambda$ varies from 1 to 10,000. Furthermore, the importance of an accurate estimate of the bandgap energy in the super-lattice structure is emphasized by assigning a value of 1 million for the coefficients $\lambda_3$ and $\lambda_4$ in equation (1).

The following segment elucidates the correlation between OIPs and the reported experimental values of the bulk optoelectronic properties. These relationships provide a framework for refining the OIPs, thereby enhancing the precision of optoelectronic property estimation at their experimental points. The generation of these equations was facilitated through the diagonalization of the Hamiltonian matrix at the Γ-point in first Brillouin zone, and incorporating the OIPs within the process. In this study, we have employed certain mathematical relationships to fulfill the condition $E_{hh} = E_{lh} = 0$. The relationship is defined as follows:

$$E_{xaxc} = \sqrt{(E_{pa} + \frac{\Delta_a}{3})(E_{pc} + \frac{\Delta_c}{3})} \quad (3)$$

This equation (3) is used to derive the value of $E_{pc}$ under the condition that $E_{so} = -\Delta$, and the resulting equation is:

$$E_{pc} = \frac{\Delta^2 + \Delta\left(E_{pa} - \frac{2}{3}\Delta_a - \frac{2}{3}\Delta_c\right) + \frac{1}{3}\Delta_a\Delta_c - E_{pa}\Delta_c}{\Delta_a - \Delta} \qquad (4)$$

Additionally, $E_{pa}$ can be computed using equation (4) as follows:

$$E_{pa} = \frac{E_{so1}(\Delta_a - \Delta) - \Delta_a\Delta + \frac{2}{3}\Delta_a^2 + \frac{1}{3}\Delta_a\Delta}{\Delta_a - \Delta_c} \qquad (5)$$

In the above equation, $E_{so1}$ denotes the energy of the spin-orbit antibonding. Furthermore, the relationship below also holds:

$$E_{sasc} = -\sqrt{E_g^2 - E_g(E_{sa} + E_{sc}) + E_{sa}E_{sc}} \qquad (6)$$

These mathematical relationships aid in reducing the number of parameters that need to be optimized. Other relationships have been mentioned in previous studies [15, 18] to estimate the energy levels at various points of the Brillouin zone, as well as in [25, 26] to estimate the curve of the graph related to the effective mass. In this research, we have used constant values for $\Delta_a$ and $\Delta_c$ for anion and cation in distinct materials, as per the guidelines in previous research [18]. These relationships are applied to each material based on the sp3s* method, considering the first nearest neighbors and the spin-orbit effect. Consequently, only *nine* OIPs are left to be optimized using the GA.

The study innovates a new methodology apt for optimizing the OIPs to accurately forecast and estimate the various properties of quantum structures, including the bulk, super-lattices, and quantum wells. Through the incorporation of super-lattice specific parameters $\lambda_3|W - W'|$ and $\lambda_4|Z - Z'|$ into the cost function, the optimization problem for OIPs is refined to simultaneously optimize both quantum structure bulk and super-lattice band gaps, overcoming the limitation of prior research which predominantly optimized OIPs only for the bulk material. The first two terms of the enhanced cost function minimize the deviation in bulk quantum structure properties, while the additional terms account for errors in super-lattice band gap energies, the $\lambda$ factors modulating contribution from both parts in the overall cost. This new cost function, embodying the total error of both the bulk and super-lattice quantum parameters prediction, can effectively direct the problem-solving process to find OIPs that, when optimized, best describe the optoelectronic properties of the entire quantum structures—be it bulk or super-lattice, thereby achieving improved parameter estimation. It's important to note that the success of this method hinges critically on the careful tuning of the lambda values $\lambda_1, \lambda_2, \lambda_3$ and $\lambda_4$ as they carry significant weight in the optimization process.

To summarize the process depicted in Figure 1, we commence by randomly initializing nine OIPs for each bulk material. Subsequently, these OIPs are subjected to Equations (3-6), yielding 15 OIPs per bulk material, which are then optimized throughout the process. The Hamiltonian matrix for bulk materials is constructed utilizing these OIPs, in conjunction with the τ, g, and K vectors according to the appendix. This process is mirrored for super-lattice structures, where the Hamiltonian matrix is formulated using 18 OIPs (9 for each bulk), alongside the corresponding matrices and vectors.

The next phase involves deriving the optoelectronic properties from the Hamiltonian matrix's eigenvalues for both bulk and super-lattice structures. For instance, to ascertain the bandgap, the wave vector K must be set to the Γ point, $K = (0,0,0)$, with the bandgap being the energy difference between the conduction and valence bands at this juncture. The selection of an appropriate K vector is crucial for other optoelectronic properties, correlating to specific points in the band structure.

Once the Hamiltonian matrix is established and its eigenvalues computed, these values facilitate the determination of various optoelectronic characteristics of the materials. These characteristics are represented as $X_i$ and $Y_i$ for bulk materials, and $W$ and $Z$ for super-lattices. For clarity, the methodology for calculating optoelectronic properties is encapsulated by Equation (2).

The process concludes with the calculation of the cost function, which is based on the discrepancy between the predicted and experimental values of the optoelectronic properties. The GA then updates the OIPs, and these revised OIPs are reutilized in an iterative process to enhance the accuracy of optoelectronic property predictions. To facilitate reproducibility and further research, the codes[1] corresponding to the methodologies and analyses described herein are available.

The optimized OIPs are used to calculate the bandgap energy in the quantum structure. However, it is important to note that the computation of an additional optoelectronic property may not yield a highly accurate estimation. Therefore, to estimate additional quantum structure properties, it becomes necessary to re-optimize the OIPs based on the new experimental data. In this study, the accurate estimation of the bandgap energy in the $(GaAs)_m/(AlAs)_n$ super-lattice is considered, and the cost function is formulated based on the experimental data of the quantum structure's bandgap energy. To accurately estimate the effective mass of quantum structures, it is necessary to optimize the OIPs using experimental data associated with the effective mass of the quantum structure in question. Moreover, both effective mass and bandgap energy can be incorporated simultaneously in the formulation of the cost function, allowing the optimized OIPs to estimate both optoelectronic properties. As a validation step for the proposed method, we plan to use a GA to optimize the OIPs for *GaAs* and *AlAs*. This will provide a more accurate estimate of the bandgap energy in quantum structures. The results of the proposed method will be compared with those of different methods in the subsequent section of this paper.

## Implementation

The following section presents the outcomes of the proposed methodology, which employs the GA. The optimized OIPs for two primary materials, *GaAs* and *AlAs*, are tabulated in Table 1. The cost function in the proposed method is formulated based on the experimental data related to the optoelectronic properties of *GaAs* and *AlAs* in their bulk states, as well as the $(GaAs)_m/(AlAs)_n$ super-lattice. This analysis is particularly focused on cases where *m=9, n=4*, and *m=10, n=4*. The necessary experimental data are provided in Table 2 and in [27]. Although it would have been possible to use experimental data from other super-lattice structures, the choice was made to limit the analysis to these specific cases. This decision was made to ensure the validation of the proposed method against the predictions of experimental data, which were not used in the estimation of the OIPs. Upon determination of the optimal OIPs, the proposed approach for evaluating the optoelectronic properties of quantum and bulk structures is depicted in Figure 2. This approach aims to provide a comprehensive understanding of the behavior and characteristics of these structures.

---

[1] https://github.com/Ali-Hajiebrahim-Zargar/OIPs-Calculation

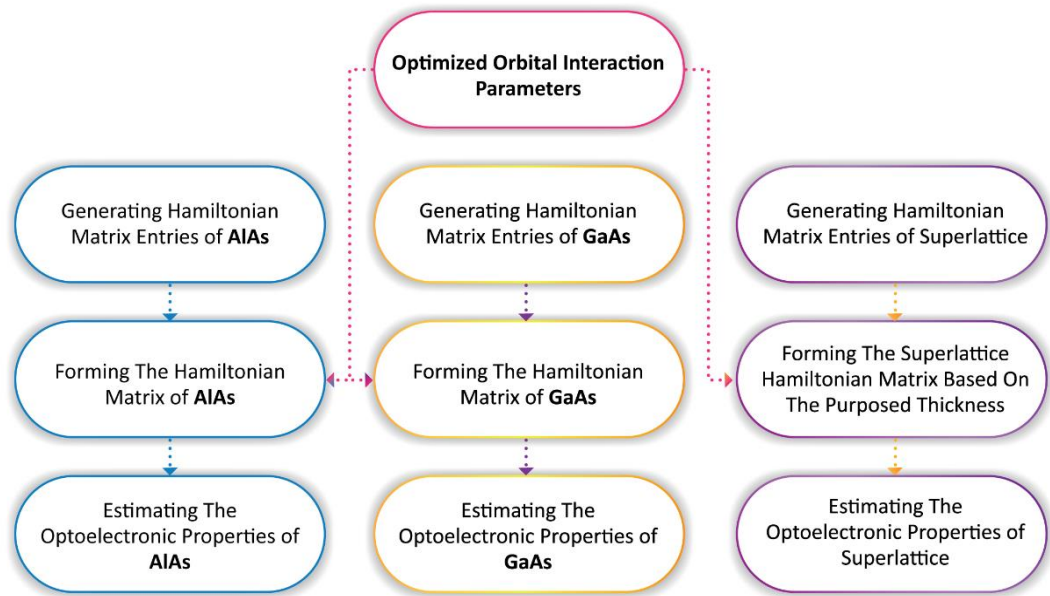

Figure 2: Schematic representation of the proposed methodology for the estimation and prediction of optoelectronic properties in quantum structures.

As delineated in Figure 2, it is evident that a substantial segment of the proposed methodology, which encompasses the calculation of Hamiltonian matrices in both the bulk and super-lattice, adheres to a uniform and systematic procedure. This procedure is consistently applied during the two key phases: parameter optimization and property estimation.

Table 1. Representation of the 15 OIPs within the Hamiltonian Matrix under the premise of sp3s* first nearest-neighbor parametrization and factoring in spin-orbit interaction, observed at a thermal field of 300 K

| OIPs | GaAs | AlAs |
|---|---|---|
| $E_{sa}(eV)$ | -4.7642 | -8.1639 |
| $E_{sc}(eV)$ | -6.0354 | -0.6369 |
| $E_{ssa}(eV)$ | 9.0528 | 14.9740 |
| $E_{ssc}(eV)$ | 5.3134 | 7.1118 |
| $E_{xayc}(eV)$ | 5.2952 | 4.6210 |
| $E_{saxc}(eV)$ | 1.9014 | 7.4231 |
| $E_{xasc}(eV)$ | 11.6705 | 6.7832 |
| $E_{ssaxc}(eV)$ | 3.8331 | 7.3042 |
| $E_{xassc}(eV)$ | 4.7758 | 3.1458 |
| $E_{pa}(eV)$ | 1.5776 | 1.4693 |
| $E_{pc}(eV)$ | 3.2967 | 3.3875 |
| $E_{sasc}(eV)$ | -6.7941 | -6.3951 |
| $E_{xaxc}(eV)$ | 2.4006 | 2.3378 |
| $\Delta_a(eV)$ | 0.421 | 0.421 |
| $\Delta_c(eV)$ | 0.174 | 0.024 |

The OIPs discussed in this study, beyond their applicability in estimating bulk optoelectronic properties, demonstrate proficiency in predicting the optoelectronic properties of various quantum structures. This includes structures such as super-lattices and quantum wells, further extending the versatility of the OIPs.

Table 2 presents a comparison between experimental and theoretical values for the optoelectronic attributes of the bulk materials, *GaAs*, and *AlAs*. This juxtaposition serves to validate the computational accuracy of our proposed method and its effectiveness in predicting the optoelectronic properties of these materials.

Table 2: Comparative Analysis of the Targeted and Calculated Properties of *GaAs* and *AlAs* Bulk Materials Using sp3s* First Nearest Neighbor Approximation and Spin-Orbit Interaction

| parameters | GaAs | | AlAs | |
|---|---|---|---|---|
| | Target | Present Work | Target | Present work |
| $\Gamma_{6c}(eV)$ | 1.424 | 1.424 | 3.020 | 3.020 |
| $\Delta_{so}(eV)$ | 0.340 | 0.340 | 0.300 | 0.300 |
| $\frac{m_\Gamma^*}{m_0}$ | 0.067 | 0.068 | 0.150 | 0.150 |
| $\frac{m_{lh}^*}{m_0}[001]$ | -0.087 | -0.076 | -0.163 | -0.154 |
| $\frac{m_{lh}^*}{m_0}[011]$ | -0.080 | -0.071 | -0.140 | -0.142 |
| $\frac{m_{lh}^*}{m_0}[111]$ | -0.079 | -0.054 | -0.135 | -0.107 |
| $\frac{m_{hh}^*}{m_0}[001]$ | -0.403 | -0.417 | -0.516 | -0.594 |
| $\frac{m_{hh}^*}{m_0}[011]$ | -0.660 | -0.658 | -1.098 | -0.907 |
| $\frac{m_{hh}^*}{m_0}[111]$ | -0.813 | -0.799 | -1.570 | -1.073 |
| $\frac{m_{so}^*}{m_0}$ | -0.150 | -0.154 | -0.240 | -0.265 |
| $L_{6c}(eV)$ | 1.708 | 1.768 | 2.352 | 3.890 |
| $\Gamma_{6v}(eV)$ | -13.100 | -12.224 | -11.950 | -11.821 |
| $\Gamma_{7c}(eV)$ | 4.530 | 4.818 | 4.540 | 4.987 |
| $\Gamma_{8c}(eV)$ | 4.716 | 5.073 | 4.690 | 5.005 |
| $X_{5v}(eV)$ | -6.800 | -5.245 | -5.690 | -6.762 |
| $X_{6v}(eV)$ | -2.880 | -2.986 | -2.410 | -2.372 |
| $X_{7v}(eV)$ | -2.800 | -2.872 | -2.410 | -2.210 |
| $X_{6c}(eV)$ | 1.980 | 1.816 | 2.229 | 3.479 |
| $X_{7c}(eV)$ | 2.320 | 3.477 | 3.800 | 6.16 |
| $L_{5v}(eV)$ | -8.000 | -5.316 | -6.000 | -6.425 |
| $L_{6v}(eV)$ | -1.420 | -1.687 | -0.880 | -1.335 |
| $L_{7v}(eV)$ | -1.200 | -1.398 | _ | -1.090 |
| $L_{7c}$ (eV) | 5.740 | 4.059 | 5.860 | 5.300 |

The experimental measurements of optoelectronic properties such as $\Gamma_{6c}, X_{6c}, X_{7v}, L_{6c}, L_{5v}, \frac{m^*_{so}}{m_0}, \frac{m^*_{lh}}{m_0}[001], \frac{m^*_{hh}}{m_0}[001]$, and $\frac{m^*_\Gamma}{m_0}$ in the bulk of *GaAs* have been previously documented in cited literature by Klimeck [20]. The proposed methodology aims to replicate these characteristic values, delivering results that are closely aligned with those presented by Klimeck [20]. However, discrepancies are noted when compared to the target values reported in another study [28]. The predictive capacity of these attributes was scrutinized using the Mean Absolute Percentage Error (MAPE) criterion as a tool for comparison across different studies. The inaccuracies in the study [28] for the *PBE*, *MBJLDA*$_{bgfit}$, and *HSE*$_{bgfit}$ methodologies are found to be 37%, 14%, and 103% respectively. In contrast, the percentage error in the proposed model and Klimeck's research [20] are significantly lower, at 7.2% and 7.6% respectively. The focus of the proposed methodology is to optimize OIPs for the accurate prediction of optoelectronic properties in quantum structures. Despite this, it has demonstrated robust performance in estimating bulk properties, showing its relative strength when compared to other methodologies.

This study aims to examine the effectiveness of the proposed methodology for determining the bandgap energy of the super-lattice structure, referred to as $(GaAs)_m/(AlAs)_n$. This is done by contrasting it with comparable methodologies that employ the TB and DFT approach. The findings of this comparative analysis are set forth in Tables 3 and 4.

Table 3: Comparative Analysis of Bandgap Energy for $(GaAs)_m/(AlAs)_n$ Super-lattices. The calculations were performed utilizing methods proposed by Klimeck et al., and OIPs from Table 1. The terms 'I' and 'D' are representative of indirect and direct bandgap respectively.

| Thickness (ML) | | Theoretical result of band gap (eV) | | | Experimental result (eV) |
|---|---|---|---|---|---|
| m | n | Present work | Vogl et. al | Klimeck et. al | PL |
| 3 | 3 | **2.02**(I) | 1.84(D) | 1.77(D) | 2.09(I)[27] |
| 5 | 5 | **1.94**(I) | 1.73(D) | 1.72(D) | 2.01(I)[27] |
| 8 | 8 | **1.83**(D) | 1.66(D) | 1.65(D) | 1.88(I)[27] |
| 6 | 3 | **1.89**(D) | 1.73(D) | 1.66(D) | 1.91(D)[28] |
| 9 | 4 | **1.75**(D) | 1.64(D) | 1.61(D) | 1.75(D)[25] |
| 10 | 4 | **1.72**(D) | 1.61(D) | 1.59(D) | 1.71(D)[25] |

The Mean Absolute Percentage Error (MAPE) values for the bandgap energy estimations of six super-lattices, as depicted in Table 3, exhibit distinct differences in the effectiveness of the methods employed by Klimeck[20], Vogl[18], and the method proposed in this study. The MAPE values corresponding to these methods are 11.67%, 9.85%, and 1.85%, respectively. This quantitative comparison underscores the superior precision of the proposed method in estimating the bandgap energy of super-lattices, relative to the other methods assessed.

Table 4 compares the outcomes of the proposed method to DFT-based methods for calculating the bandgap energy of super-lattices.

Table 4: A Comparative Analysis of Bandgap Energy Estimations in Super-lattice Structures of *GaAs* and *AlAs* Using the Proposed Method and DFT-Based Approaches.

| Thickness (ML) | | Theoretical result of band gap (eV) | | | | Experimental result (eV) |
|---|---|---|---|---|---|---|
| n | m | *DFT* | DFT with scissor correction | Hybrid DFT | Present work | PL |
| 1 | 1 | 1.14 | 2.04 | 2.06 | **2.07** | 2.07[31] |
| 2 | 2 | 1.08 | 1.98 | **1.99** | 1.985 | 2.097[32] |
| 3 | 3 | 1.00 | 1.90 | 1.93 | **2.02** | 2.09[29] |

The data presented in the above table reveals the superior precision of the proposed method in estimating bandgap energy, as compared DFT-based techniques, which are typically employed for bandgap energy corrections.

This methodology can be further extended to predict bandgap energy in quantum wells, specifically in ternary alloys, by utilizing optimized OIPs. For evaluation purposes, the quantum well structure $Al_xGa_{1-x}As/GaAs$ is considered. In this context, GaAs serves as the quantum well. Subsequently, the Hamiltonian matrix is constructed, and Wigard's law is implemented to optimize the OIPs based on the composition of each material. It is important to note that modern estimates of ternary structures integrate additional methods, as detailed in [33], which may be taken into account during the calculation of OIP values for such structures.

To validate the accuracy of the bandgap energy estimation in $Al_xGa_{1-x}As/GaAs$, three samples, namely C165, C166, and C167, as referenced in [34], are examined. These samples were cultivated at a temperature of 300K on a *GaAs* wafer(100). The results obtained using the enhanced OIPs, as displayed in Table 1, are presented in Table 5, alongside the findings of experimental value.

Table 5: Comparative Analysis of Experimental and Theoretical Bandgap Energies for C165, C166, and C167 via the Implemented Technique.

| *Ternary Alloy* | *Present work (eV)* | *Experimental result (eV)* |
|---|---|---|
| C165 | 1.45 | 1.47 |
| C166 | 1.49 | 1.54 |
| C167 | 1.47 | 1.48 |

The results derived from the proposed methodologies for quantum wells, specifically those based on $Al_xGa_{1-x}As/GaAs$, exhibit a considerable similarity to the empirically obtained values for bandgap energy.

Further extensions of this study are discussed subsequently in this section. In particular, this research employed optimized OIPs to predict the energy gaps of quantum structures, such as super-lattices and quantum wells, in cases where experimental values were not readily available. To illustrate this, we present a graph depicting the variations in the cutoff wavelength of the *GaAs* quantum well in $Al_xGa_{1-x}As/GaAs$. This graph, shown in Figure 3, details these variations according to various compositions and optimized OIPs.

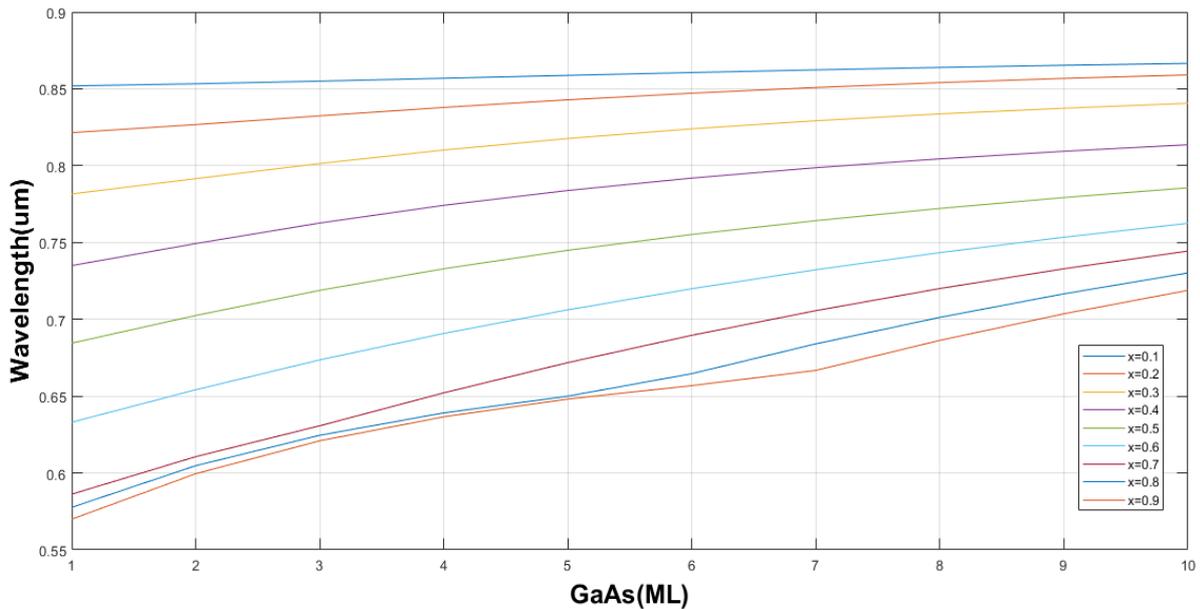

Figure 3: The correlation between cut-off wavelength (μm) and *GaAs* thickness in a mono-layer, considering different *Al* composition in $Al_xGa_{1-x}As$.

Based on Figure 3, it is observed that increasing the thickness of the *GaAs* layer in $Al_xGa_{1-x}As/GaAs$ quantum well results in a decrease in the bandgap, leading to an expected increase in the cutoff wavelength. Additionally, there is an observed increase in the cutoff wavelength with changes in the composition of *Al*. It is known that the bandgap of *AlAs* is significantly larger than that of *GaAs*. Therefore, as the proportion of *Al* in the $Al_xGa_{1-x}As$ alloy increases, the bandgap of the alloy approaches that of *AlAs*. This leads to an increase in the bandgap of the quantum well structure. Consequently, the cutoff wavelength decreases with an increase in the *Al* composition, reflecting the relationship between the bandgap and the cutoff wavelength in these semiconductor materials.

Moreover, it's worth mentioning that, this plot is derived by altering the thickness of the *GaAs* layer and the *Al* composition in the Hamiltonian matrix. Optimized OIPs, as outlined in Table 1, are first computed for the alloy structure $Al_xGa_{1-x}$ using Vegard's Law. These parameters are then passed into the Hamiltonian matrix. The wave vector K is set based on the location of the minimum of the conduction band, which is essential for constructing the Hamiltonian matrix. Subsequently, by analyzing the eigenvalues of this matrix, the bandgap and cutoff wavelength of the material can be determined.

## Conclusion

In this research, we introduce a novel method based on empirical TB theory for the estimation of quantum structures' properties, as outlined in the following diagram.

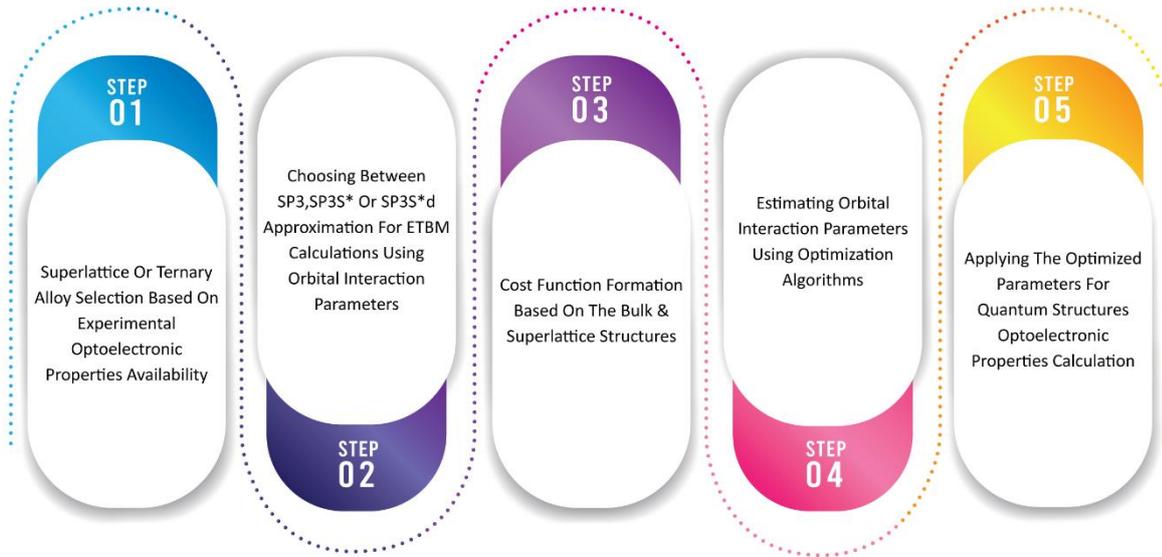

Figure 4: Outline of the proposed method

The initial phase involves the selection of quantum structures for which experimental optoelectronic property values are available for both the bulk and super-lattice. Subsequently, the optoelectronic properties as functions of the OIPs are calculated, necessitating an appropriate approximation for TB-based computations (refer to equation 2). Following this, a cost function specific to the quantum structure is established, which should integrate four components as denoted in equation 1. The first two terms correspond to the cost function for the bulk of the quantum structures, whereas the third and fourth terms represent two instances of super-lattices, quantum wells, or a combination thereof, for which the existence of the experimental bandgap energy is recognized. The OIPs are subsequently computed and optimized through optimization algorithms, based on the mathematical relationships in the cost function (refer to equations 3-6). In the terminal phase, these parameters are utilized to estimate or predict the bandgap energy and other optoelectronic properties of other relevant quantum structures.

The method proposed in this study was evaluated using the quantum structure of *GaAs/AlAs*. The results demonstrated that the proposed method for estimating the energy gap of the $(GaAs)_m/(AlAs)_n$ super-lattice surpasses previous methods such as TB and DFT-based approaches. Additionally, this method exhibited efficacy in estimating the bandgap energy of f $Al_xGa_{1-x}As/GaAs$ quantum wells, which was in close agreement with experimental values.

## Future work

In light of the findings from this study, the cost function for the super-lattice structure was derived solely based on the bandgap energy. Given that there are limited experimental data available on the optoelectronic properties of super-lattice structures, DFT and band structure simulations can be employed as a valuable resource. The data points obtained from these computational methods could potentially serve as substitutes for experimental data points in the proposed methodology.

Moreover, implementing more precise approximations in the TB, such as the sp3ds* approximation, could potentially enhance the proposed method's capacity to estimate optoelectronic and transmission properties. Such an approach could provide a more comprehensive understanding of the super-lattice structure's behavior and improve the accuracy of the predictions.

Furthermore, the proposed methodology has potential applications beyond the scope of this study. For instance, it could be employed to predict the optoelectronic properties of structures with varying dimensions, such as one-dimensional Quantum Wire structures and zero-dimensional Quantum Dot structures. This could open up new avenues for research and development in the field of optoelectronics, thereby contributing to the advancement of this scientific field.

## Appendix: Bulk Hamiltonian Matrix

$$H_{Bulk} = \begin{bmatrix} E_{sa} & 0 & 0 & 0 & 0 & g_0 E_{sasc} & g_1 E_{saxc} & g_2 E_{saxc} & g_3 E_{saxc} & 0 \\ 0 & E_{pa} & -i\frac{\Delta_a}{3} & \frac{\Delta_a}{3} & 0 & -g_1 E_{xasc} & g_0 E_{xaxc} & g_3 E_{xayc} & g_2 E_{xayc} & -g_1 E_{xassc} \\ 0 & i\frac{\Delta_a}{3} & E_{pa} & -i\frac{\Delta_a}{3} & 0 & -g_2 E_{xasc} & g_3 E_{xayc} & g_0 E_{xaxc} & g_1 E_{xayc} & -g_2 E_{xassc} \\ 0 & \frac{\Delta_a}{3} & i\frac{\Delta_a}{3} & E_{pa} & 0 & -g_3 E_{xasc} & g_2 E_{xayc} & g_1 E_{xayc} & g_0 E_{xaxc} & -g_3 E_{xassc} \\ 0 & 0 & 0 & 0 & E_{ssa} & 0 & g_1 E_{ssaxc} & g_2 E_{ssaxc} & g_3 E_{ssaxc} & 0 \\ g_0^* E_{sasc} & -g_1^* E_{xasc} & -g_2^* E_{xasc} & -g_3^* E_{xasc} & 0 & E_{sc} & 0 & 0 & 0 & 0 \\ g_1^* E_{saxc} & g_0^* E_{xaxc} & g_3^* E_{xayc} & g_2^* E_{xayc} & g_1^* E_{ssaxc} & 0 & E_{pc} & -i\frac{\Delta_c}{3} & \frac{\Delta_c}{3} & 0 \\ g_2^* E_{saxc} & g_3^* E_{xayc} & g_0^* E_{xaxc} & g_1^* E_{xayc} & g_2^* E_{ssaxc} & 0 & i\frac{\Delta_c}{3} & E_{pc} & -i\frac{\Delta_c}{3} & 0 \\ g_3^* E_{saxc} & g_2^* E_{xayc} & g_1^* E_{xayc} & g_0^* E_{xaxc} & g_3^* E_{ssaxc} & 0 & \frac{\Delta_c}{3} & i\frac{\Delta_c}{3} & E_{pc} & 0 \\ 0 & -g_1^* E_{xassc} & -g_2^* E_{xassc} & -g_3^* E_{xassc} & 0 & 0 & 0 & 0 & 0 & E_{ssc} \end{bmatrix}$$

The asterisk (*) denotes the complex conjugate operation. The coefficients $g_0$, $g_1$, $g_2$, and $g_3$ are defined as follows:

$$g_0 \equiv \frac{\left[\exp(i\vec{k}.\vec{\tau_1}) + \exp(i\vec{k}.\vec{\tau_2}) + \exp(i\vec{k}.\vec{\tau_3}) + \exp(i\vec{k}.\vec{\tau_4})\right]}{4}$$

$$g_1 \equiv \frac{\left[\exp(i\vec{k}.\vec{\tau_1}) - \exp(i\vec{k}.\vec{\tau_2}) + \exp(i\vec{k}.\vec{\tau_3}) - \exp(i\vec{k}.\vec{\tau_4})\right]}{4}$$

$$g_2 \equiv \frac{\left[\exp(i\vec{k}.\vec{\tau_1}) - \exp(i\vec{k}.\vec{\tau_2}) - \exp(i\vec{k}.\vec{\tau_3}) + \exp(i\vec{k}.\vec{\tau_4})\right]}{4}$$

$$g_3 \equiv \frac{\left[\exp(i\vec{k}.\vec{\tau_1}) + \exp(i\vec{k}.\vec{\tau_2}) - \exp(i\vec{k}.\vec{\tau_3}) - \exp(i\vec{k}.\vec{\tau_4})\right]}{4}$$

The vectors $\vec{\tau_1}, \vec{\tau_2}, \vec{\tau_3}, \vec{\tau_4}$ are determined based on the lattice shape, with $k$ being the reciprocal lattice vector. For the Zinc Blende structure, these vectors are defined as:

$$\vec{\tau_1} = \frac{a}{4}(1,1,1)$$
$$\vec{\tau_2} = \frac{a}{4}(-1,-1,1)$$
$$\vec{\tau_3} = \frac{a}{4}(1,-1,-1)$$
$$\vec{\tau_4} = \frac{a}{4}(-1,1,-1)$$